\documentstyle[editedvolume]{crckapb}

\def \o{\omega}
\def \a{\alpha}
\def \t{\tilde}
\def \i{\int}
\def \p{\partial}
\def \mpl{m_{pl}^2}
\def \s{\sigma}
\def \n{\nonumber}
\def \d{\delta}
\def \m{g^{\mu\nu}}
\def \rg{(-g)^{1/2}}
\def \e{\epsilon}
\def \G{\Gamma}
\def \L{\Lambda}
\def \l{\lambda}

\def \be{\begin{eqnarray}}
\def \en{\end{eqnarray}}

\begin{opening}
\title{INFLATIONARY UNIVERSE IN\protect\\
       KALUZA-KLEIN THEORIES}

\author{A. S. Majumdar}

\institute{S.N.Bose National Centre for Basic Sciences\\
Block JD, Sector III, Salt Lake, Calcutta 700091, India}

\end{opening}

\begin{document}

We describe extended inflation and its typical problems. We then briefly
review essential features of Kaluza-Klein theory, and show that it leads
to a scenario of inflationary cosmology in four dimensions. The problem
of stable compactification of extra spatial dimensions is discussed. The
requirements for successful exetended inflation lead to constraints on
the parameters of higher dimensional models.

\section{Introduction}

The standard big-bang model~[1] describing a homogeneous and isotropic
universe with Freidmann-Robertson-Walker (FRW) metric, is to date, the
most successful model of cosmology. However, it is well known that the
standard cosmological model is plagued with certain key problems like
the horizon, flatness, entropy, and structure formation problems~[2].
 The most plausible solution to these and other  problems
of this  model is provided by the inflationary scenario~[3].
Inflation entails a period of expansion of the cosmic scale
factor by at least a factor of $10^{28}$. The widely accepted
mechanism to drive this large expansion is through
the means of a scalar field stuck in a local minimum of its
potential with a huge energy density. The `old' inflationary
model~[4] is based on a first order phase transition of a scalar
field, as it tunnels through the potential barrier towards the
absolute minimum. The transition proceeds via nucleation of
bubbles of the true vacuum phase. During this process the dominant
vacuum energy of the scalar field drives an exponential expansion
of the scale factor. The bubbles of the new phase are unable
to meet in an expanding background and hence, the true vacuum
never percolates. Thus one faces the ``graceful exit problem''~[5].

 A variant of this scenario, i.e., the so called `new' inflationary
model~[6], relies on a slow rollover phase transition of the
scalar field to solve the graceful exit problem. However, here
one requires the effective potential to be extremely flat near
the origin. This necessitates large fine-tuning of the potential
parameters. Similar fine-tuning problems are also encountered
in `chaotic' inflation~[7].

 Extended inflation (EI) [8] restores the spirit of `old' inflation
in the sense that the scalar field herein undergoes a strongly
first order phase transition. However, here the Einstein theory
of gravity is replaced in general by scalar-tensor theories
[9]. The simplest such theory is the Jordan-Brans-Dicke (JBD)
theory [10], which was used by La and Steinhardt [8] 
for obtaining a model of EI which could remove the graceful exit
problem of `old' inflation. However, it was soon realized that this
simple model of EI had some typical problems of its own, and the search
continued for better models. 

In this review we shall first set up the framework used for EI in
context of the simple JBD model. In doing so, we shall see how certain
basic problems crop up in implementing EI (Section 2). As we shall
note, a viable EI scenario is naturally incorporated for the 4-dimensional
spacetime of higher dimensional (Kaluza-Klein) (KK) theories. In 
Section 3 we shall briefly introduce the rudimentary features of KK
theories before showing how inflation is obtained for the 4-dimensional
spacetime of one simple KK model. An essential feature of KK theories
is that the extra spatial dimensions are compactified to a tiny radius.
We shall see how this occurs for the case of more sophisticated models
in Section 4. The requirement of obtaining viable density perturbations
and observations on the cosmic microwave background radiation (CMBR) 
places severe restrictions on the potential of the scalar field used in
inflation. We conclude by describing the constraints imposed on the
parameters of KK models by the conditions for successful EI in Section 5.

\section{Extended Inflation and its problems}

Extended inflation was first implemented~[8] using a simple JBD model
coupled to matter fields, having an action given by
\be
S = \int d^4x (-g)^{1/2}[-\phi R + \o g^{\mu\nu} 
{\p_{\mu}\phi\p_{\nu}\phi \over \phi} + 16\pi {\cal L}_{matter}]
\en
The JBD field $\phi$ plays the role of a time varying gravitational
constant. $\omega$ is a free parameter called the JBD parameter.
The Einstein theory follows in the limit $\omega \to \infty$
[9]. The present day accuracy of time delay experiments
require $\omega > 500$ 
[11]. The matter sector ${\cal L}_{matter}$ is assumed to contain a scalar
field $\chi$ which undergoes a first order phase transition from a
metastable state of its potential (false vacuum) to the energetically
favorable absolute minimum (true vacuum). The
spacetime geometry is described by a Freidmann-Robertson-Walker
(FRW) metric 
\be
ds^2 = dt^2 - a^2(t)[{dr^2 \over 1-kr^2} + r^2 d\Omega^2]
\en
 where $a(t)$ is the FRW scale factor and $k = -1, 0, +1$ denotes
an open, flat and closed geometry respectively. The equations
of motion that follow from (1) are
\be
{\dot{a}^2 \over a^2} = {8\pi\rho \over 3\phi} - {k \over a^2}
+ {\o \dot{\phi}^2 \over 6\phi^2} - {\dot{a}\dot{\phi} \over a\phi} \n \\
\ddot{\phi} + {3\dot{a}\dot{\phi} \over a} = {8\pi(\rho - 3p) \over 3 + 2\o}
\en
 where $\rho$ and $p$ are the energy density and pressure
of matter respectively. During the inflationary era the energy
density is dominated by the vacuum energy of the inflaton field
and the equation of state is given by $p = - \rho$. The
term $k/a$ is insignificantly small and the following
solutions are obtained.
\be
\phi(t) = \mpl(1 + {c_1t \over c_2})^2 ~;~
a(t) = (1 + {c_1t \over c_2})^{\o + 1/2}
\en 
with $c_1^2 = (8\pi\rho)/3\mpl$ and $c_2^2  = (3+2\o)(5+6\o)/12$.
 Here $\mpl$ is a constant of integration
corresponding to the effective Planck mass at the beginning
of inflation $(t=0)$. The actual Planck mass should correspond
to the current value of $\phi$~[28]. ($G^{-1}=\mpl$, in units of $c=\hbar =1$). 

In the limit of $\o \to \infty$ equations (4) give
a constant solution $\phi = \mpl$,
and an exponential solution for the scale factor $a(t)$,
thus reproducing the `old' inflationary scenario~[3]. However,
for small enough $\o$ defined by the condition, $c_1t > c_2$,
the solutions take the shape
\be
\phi(t) \approx \mpl ({c_1t \over c_2})^2 ~;~
a(t) \approx ({c_1t \over c_2})^{\o + 1/2}
\en

This power law expansion of the scale factor continues till
the completion of the phase transition of the inflaton field.
Bubbles of true vacuum are produced and their nucleation rate
per unit volume can be expressed as 
$\G = Aexp(-B)$
where $A$ is a constant with units of (mass) and
$B$ is the action corresponding to the least resistance path across
the potential barrier. This action is called the bounce action~[12].
 Taking into account the expansion of the universe, the
stage of progress of the first order phase transition is determined
by the parameter $\e(t)$ given by
\be
\e(t) = {\G \over H^4(t)}
\en
 where  $H(t) \equiv \dot{a}(t)/a(t)$.
$\e$ reaching order unity denotes completion of the phase
transition. Nevertheless, $\e$ should be small enough
initially to allow for more than 66 e-foldings $(e^{66} \sim 10^{28})$ of
the scale factor necessary to solve the horizon problem. The
nucleation rate $\G$ depends upon the parameters of the
inflaton potential and is a constant here. For a power law expansion,
$H$ goes as $1/t$ and hence $\e(t)$ keeps on increasing continuously
to reach order unity, at which point the universe is dominated
by bubbles of true vacuum. These bubbles collide with each other,
with the energy in their walls being converted into the thermal
energy of particles and radiation. The universe is reheated
and henceforth its evolution proceeds as in a standard radiation
dominated era. The JBD field $\phi$ keeps on growing slowly
to reach its current value of $\mpl$.

 Although the percolation of true vacuum is possible, this scenario
leads to a nearly flat bubble distribution [13]. This means
that at any given instant, bubbles with all sorts of sizes are
present. This can hamper efficient thermalization. Also, the
 initially produced bubbles can grow out to large sizes, which
can later create unwanted inhomogeneities in the CMBR. The fraction
of volume occupied by bubbles greater than a certain radius
$r$, at the end of the phase transition $(t_{end})$
is given by [13]
\be
f(x > r, t_{end}) \approx (H.x)^{-4/\o}
\en
For $H.x \sim 10^{28}$,  $f$ should
be less than $10^{-4}$ to satisfy the CMBR
 constraints.  This immediately leads to a bound on $\o$, i.e.,
$\o < 20$. But as mentioned earlier, JBD theory
is consistent with observations only if $\o > 500$. This
contradiction is called the $\o$ problem.

 Several ways have since been suggested to save extended inflation.
Among these are variable $\o$ theories [14], wherein
$\o$ is a function of the JBD field as incorporated in
more general scalar-tensor theories [9]. However, the motivation
needed to consider such theories is not clear. Modification
of the inflaton sector has been considered by certain authors
[14,15], with the aim of making the nucleation rate $\G$ time
dependent, thereby lifting the upper bound on $\o$ to
an acceptable value. Several efforts have gone in the direction
of building JBD models with ad hoc potentials for $\phi$ [16].
The philosophy behind such constructions is to allow $\phi$
to be anchored by these potentials after completion of
the inflationary phase transition. In this way an effective
Einstein theory would emerge, eliminating the need for the lower
bound on $\o$, i.e., $\o > 500$. $\o$ can be
as small as required to facilitate the completion of phase transition
together with a bubble distribution allowed by the CMBR constraints
[13].

 The sort of scheme mentioned last seems promising enough if
implemented in a more natural framework, i,e., without having to include
ad hoc potentials. In the next sections
we shall see that dimensional reduction of several Kaluza-Klein
models give rise to JBD type actions in four dimensions with
potential terms for the JBD field $\phi$.

\section{Kaluza-Klein theory and inflation in four dimensions}

 The study of physics in more than four dimensions  received
a great boost  with the phenomenal success of string
theory. Implementation of the concept of unification has opened
up the exciting possibility that our world might consist of
extra spatial dimensions. The implications of this for cosmology
are profound. For the present context, we would like to see how a scenario
of inflationary cosmology for the 4-dimensional spacetime which can get
rid of the problems of extended inflation, mentioned earlier, can be
obtained. Further, 
it needs to be explained how there should be an
enormous difference in size between the observed four-dimensional
spacetime, and the unobservable tightly curled up extra dimensions.

 With this aim in mind, we proceed to study inflationary
cosmology in higher dimensional theories as seen from the four
dimensional point of view. To do this, one needs to establish
the correspondence of a Kaluza-Klein theory~[17,18] to the language
of the standard four dimensional field theory. More precisely,
the meaning of the various field operators in a dimensionally
 reduced field	theory has to be brought out clearly~[19]. To
illustrate this point, let us consider the simple example of
a massless scalar field in 4+D dimensions~[20]. The corresponding
wave equation can be written down as
\be
\Box_{4+D}\phi \equiv \Box_4\phi + \nabla_D\phi = 0
\en
 where $\Box_4$ denotes the d'Alembertian
in four dimensions and $\nabla_D$ the Laplacian in
$D$ dimensions (the additional dimensions are all assumed to
be spacelike). Assuming the D-dimensional manifold to be a closed
one, we can now expand $\phi$ in terms of the eigenfunctions
of the Laplacian $\nabla_D$:
\be
\phi(x,y) = \sum_{\nu} u_{\nu}(y)\phi_{\nu}(x)
\en
 where $x$ and $y$ represent the 4-dimensional and D-dimensional
coordinates respectively. $u_{\nu}$ is the $\nu$-th
eigenfunction of $\nabla_D$ with eigenvalue $\l_{\nu}$
\be
\nabla_D u_{\nu} = \lambda_{\nu}u_{\nu}
\en

 The fields $\phi_{\nu}(x)$ depend only upon the coordinates
of the four dimensional spacetime and obey the wave equation
given by (from  (8),(9),(10))
\be
(\Box_4 + \l_{\nu})\phi_{\nu} = 0
\en
 Thus, $\l_{\nu}$ will be observed as masses of the 4-dimensional
fields $\phi_{\nu}$. These masses are typically of the
order $1/b$ [19,20], where $b$ represents the size of the compact
manifold. Every single massless field in $4+D$ dimensions therefore
appears as a tower of massive states plus some massless modes
given by $\l_{\nu} = 0$. If $b$ is of the order of planck
length, the massive states will not play any role in the dynamics
and can be integrated out of the low energy effective field
theory. Similar results also hold true for particles of any
spin and mass [17-20]. This procedure of dimensional reduction
is thus applied by the harmonic expansion of higher dimensional
fields on the compact manifold, to obtain the 4-dimensional
effective field theory, retaining only the zero modes.

 We can now proceed to analyze our first KK model.
This model was considered in detail by Holman et al [21] and
shall be used by us to set up the problem of extended inflation
in a proper perspective. The action in $4+D$ dimensions is given
by
\be
\t{S} = \i d^{4+D}z(-\t{g})^{(1/2)}[-{1 \over 16\pi\t{G}}\t{R} +
{1\over 2} \t{g}^{MN}\p_M \t{\chi} \p_N \t{\chi} - \t{U}(\t{\chi})]
\en
We have used $z$ to represent the coordinates of the full $4+D$
dimensional spacetime and tildes to mark the various $4+D$ dimensional
objects. Upper case Latin indices denote the full $4+D$ dimensional 
coordinates, whereas, Greek indices denote 4-dimensional coordinates.
The scalar
field $\t{\chi}$ plays the role of the inflaton, 
and it will be assumed to be caught in a metastable state
of its potential $\t{U}(\t{\chi})$, from where it can
tunnel out by the nucleation of bubbles of the true vacuum.
The $4+D$ dimensional line element is assumed to take the form
\be
d\t{s}^2 = dt^2 - a^2(t)d\Omega_3^2 - b^2(t)d\Omega_D^2
\en
with $d\Omega_3^2$ corresponding to the
line element of a maximally symmetric 3-space with
scale factor $a(t)$ and $d\Omega_D^2$ to
that of a D-sphere with scale factor $b(t)$. As discussed earlier,
the zeroth mode in the harmonic expansion of $\t{\chi}$
is independant of the coordinates of the D-sphere. This enables
one to rewrite $\t{S}$ as
\be
\t{S} = [\i d^Dy (g_D(y))^{1/2}]S
\en
 where $g_D(y)$  is the determinant of the D-dimensional
metric. $S$ is the effective four dimensional action which is
independent of the D-coordinates represented by $y$. $S$ is given
by
\be
S = \i d^4x (-g)^{1/2} \Omega_D b^D(t)[- {R \over 16\pi \t{G}} \n \\
 -
{D(D-1) \over 16\pi \t{G}b^2}(\m \p_{\mu}b\p_{\nu}b - 1) +
{1\over 2}\m \p_{\mu}\chi \p_{\nu}\chi - U(\chi)]
\en
$R$ is the curvature scalar for the four dimensional spacetime
with metric $g_{\mu\nu}(x)$ and $\Omega_D = (2\pi^{(D+1)/2})/\G(D+1)/2$.
 Note here
that the kinetic term for the $b$-field has appeared with a negative
sign. It is not apparent that the action of (15) resembles
a JBD action. To see that, it is necessary to make the following
redefinitions:
\be
{\Omega_D b_0^D \over 16\pi \t{G}} \equiv {1 \over 16\pi G} ~;~
\s \equiv (\Omega_D b_0^D)^{1/2}\chi \n \\
V(\s) \equiv (\Omega_D b_0^D)U(\chi) ~;~
\phi \equiv {(b/b_0)^D \over 16\pi G} 
\en
 $b_0$ is a constant which sets the scale of compactification.
 $G$ is the four dimensional
Newton constant. The field $\s$ shall henceforth be called
the inflaton field. $\phi$ plays the role of a JBD
field when we rewrite $S$ as
\be
S = \i d^4x \rg [-\phi R -\o\m{\p_{\mu}\phi \p_{\nu}\phi \over \phi} \n \\
+
\a \phi^{1-2/D} + 16\pi G\phi ({1\over 2}\m\p_{\mu}\s \p_{\nu}\s -V(\s))]
\en
where, 
\be
\a = D(D-1)(16\pi G)^{-2/D}b_0^{-2} ~;~
\o = 1 - 1/D
\en

It is important to note several facts about the JBD type action
of equation (17). First, as mentioned earlier, the kinetic
energy term for the field $\phi$ occurs with a negative
sign in contrast to the usual JBD theories. Secondly, a potential
 for $\phi$ (third term)
has naturally followed. Thirdly,
$\phi - \s$ cross terms are also present. This
is an outcrop of dimensional reduction from higher dimensions
and as such has no analogue in JBD lagrangians written down
directly in four dimensions.

 At this stage it is essential to look into the dynamics of this
system to check whether extended inflation works here. As usual,
it is assumed that the inflaton field $\s$ makes a first
order phase transition from its metastable vacuum at $\s = 0$
 to the true vacuum at $\s = \s_0$.
Defining the value of the potential at the false vacuum as 
$V_0 = 8\pi G V(\s = 0)$, we have the
following equations of motion from (17):
\be
{\dot{a}^2 \over a^2} + {\dot{a}\dot{\phi} \over a\phi} +
{\o \dot{\phi}^2 \over 6\phi^2} + {\a \phi^{1-2/D} \over 6} - {V_0 \over 3}
= 0 \n \\
\ddot{\phi} + {3\dot{a}\dot{\phi} \over a} = - {dW(\phi) \over d\phi}
\en
with the potential $W(\phi)$ given by
\be
W(\phi) = {\a \phi^{2(1-1/D)} \over 2(1-1/D)} - {V_0 \phi^2 \over 1+2/D}
\en

 Before attempting to solve these equations for $a(t)$ and $\phi(t)$,
it is better to briefly recall our purpose here. We would like
to obtain expanding or inflationary solutions for the 4-dimensional
scale factor $a(t)$. Also since $\phi(t)$ is proportional
to the D-dimensional scale factor $b(t)$, we would want decreasing
solutions for $\phi(t)$. This would signify contraction
of the extra space. For successful percolation of the true phase,
we desire the inflaton phase transition to take place in a suitable
manner as stated in section 1.

The potential $W(\phi)$ (20) is dominated by the
second term for large values of $\phi$. 
It has
a maximum at $\phi_0 = ((D+2)/2D)^{D/2}$. 
If the initial value of $\phi$ is greater
than $\phi_0$, then the field $\phi$ will
roll down the potential hill towards larger values of $\phi$,
signifying expansion of the D-dimensional space. If $\phi$ starts
out with a value lesser than $\phi_0$, a simple
numerical integration of equations (19) shows that enough
expansion of the scale factor $a(t)$ is not possible, before $\phi$
is driven down to zero~[21]. 
 Hence, simultaneous inflation of the 4-dimensional
spacetime and compactification of the extra space is not possible
in this model.

Notwithstanding the fact that this model does not work, it is
interesting to note its one successful feature, namely the nature
of phase transition of the inflaton field $\s$. It is
well appreciated that the calculation of the nucleation rate
of true vacuum bubbles is quite complicated in models containing
more than one time-varying field. The present case is one
such example since here the phase transition of $\s$ takes
place in the background of an evolving field $\phi$. Due
to the $\phi - \s$ cross terms in (17),
the nucleation rate becomes time dependent. The non-flat nature
of the background spacetime complicates the matter further.
However, it was observed in refs.[22], that in the limit
of weak gravity it is possible to systematically ``freeze-out''
the gravitational effects in the bounce action. For theories
derivable from an action of the type given in (17), they
obtained a closed form expression for the nucleation rate
\be
\G = A_0(16\pi G\phi)^2 exp(-16\pi G \phi B_0)
\en
with $B_0$ the bounce action as calculated in flat
space and $A$ is proportional to $\s_0^4$.
It should be noted that the approximation of weak
gravity is valid only for large $\phi$.

The major difference of this theory from 4-dimensional EI models
is brought out by noting that both the numerator and denominator
in the expression $\e = \G/H^4$
are now time-dependent. Any solution of this theory which
proceeds towards compactification of the extra space (solutions
for $\phi$ decreasing with time), would also contribute
towards enhancing the rate of true vacuum percolation. (It can
be seen that $\phi$ is present in the exponent of the expression
for $G$ (21)). This would make it easier for
$\e(t)$ to start out from very small values and still
reach order unity to establish  completion of the phase transition.

\section{Stable compactification and successful inflation}

We have seen in the previous section that a simple model of Einstein gravity together
with an inflaton field in higher dimensions, does not produce
all the features necessary for a viable model of EI in four
dimensions. The difficulty can be summed up as a lack of compatibility
of obtaining sufficient inflation along with compactification
of the extra manifold. A natural extension is to consider a
higher dimensional model with a nonminimally coupled inflaton
field. The relevance of nonminimal coupling for scalar fields
in Kaluza-Klein theories has been observed by Sunahara et al
[23]. They found that certain range of values of the coupling
parameter could automatically prohibit the isotropic expansion
of all the 4+D dimensions. 

Let us now consider a model in ten dimensions with a nonminimally coupled
scalar field having an action given by [24]
\be
\t{S} = \i d^{10}z (- \t{g})^{1/2}[- {\t{R} \over 16\pi \t{G}} +
{1\over 2} \t{g}^{MN} \p_M \t{\chi}\p_N \t{\chi} - \xi \t{R}(\t{\chi}^2 -
\t{\chi}_0^2) - \t{U}(\t{\chi})]
\en
 Here $\xi$ is an arbitrary nonminimal coupling parameter.
One can follow the same procedure of dimensional reduction
as described in the previous section. After performing the transformations
defined in (16), the effective 4-dimensional action is obtained.

The equations of motion for the 4-dimensional scale factor $a(t)$, as well
as the 4-dimensional JBD field $\phi(t)$ (related to the internal scale
factor $b(t)$) have been derived in [24]. The potential for the JBD field
is given by
\be
W(Y)  = {3Y^{5/3} \over 5} - {3Y^2 \over 4\d}
\en
where
\be
Y = ({\a \over V_0})^{-1/3}\phi(t) ~;~
\d = 1 - 16\pi G \xi \s_0^2
\en

The equations of motion have been numerically integrated 
for negative values of $\d$. Rapid expansion of the scale
factor $a(t)$ ensues. Analysis of the
potential $W(Y)$ suggests that the initial value of $Y$ can be
as large as required, to enable completion of greater than 65
e-foldings of the scale factor $a(t)$, before $Y$ is driven down to
zero. Numerical integration confirms that one has a scenario
of rapid expansion for the 4-dimensional scale factor  together
with a simultaneous contraction of the extra dimensional scale
factor. 

 A prerequisite for a successful theory of EI, is the suitable
completion of the inflationary phase transition. To see that
this indeed is the case here, one has to calculate $\e(t)$
for various stages of the evolution of $a(t)$ and $\phi(t)$.
The results show~[24] that for a wide choice of initial nucleation
rates, it is possible to complete the phase transition before
$\phi$ becomes very small to violate the weak gravity approximation~[22].

 In this sort of a scenario the possibility of finding large-sized
leftover bubbles after inflation is extremely rare. The bubble
distribution is expected to be highly peaked around a certain
critical size, corresponding to the formation of a very large
majority of bubbles towards the end of the phase transition
when the value of $\e$ is near 1. Consequently,
the thermalization of energy in the bubble walls  proceeds
in an efficient way. It needs to be emphasized that that this
is a generic feature of these kinds of higher dimensional models,
and therefore is insensitive to the choice of various free parameters.

Thus in this model, the inclusion of a nonminimally coupled inflaton
field in higher dimensions enables the dynamical contraction
of the extra space to occur together with extended inflation
in the 4-dimensional spacetime. No upper bound on $\o$ is
needed in this model to produce a well-behaved bubble distribution.
Nevertheless, questions regarding compactification that  still remain are: 
(i) There is nothing in the theory to prevent
the shrinking of the extra dimensions to zero. (ii) One has to ensure that
decompactifying solutions for the internal scale factor $b(t) (\sim \phi(t))$
are forbidden, even after the inflaton field has settled down at its true
vacuum (i.e., $V_0 = 0$). We shall now see how these problems are tackled
by using a different KK model in ten dimensions.

To this end, a ten dimensional JBD model was considered in [25], which
has the action given by
\be
\t{S} = {1\over 16\pi}\i d^{4+D}z\rg [- \t{\Phi}\t{R} +
\t{\o}\t{g}^{MN}{\p_M \t{\Phi} \p_N \t{\Phi} \over \t{\Phi}} -
\t{\L} + \t{\cal L}(\t{\psi})]
\en
 where, $\t{\cal L}(\t{\psi})$  denotes the Lagrangian
of the inflaton field.
$\t{\L}$ is a cosmological
constant and $\t{\o}$ is the 10-dimensional JBD parameter. 
 As seen in the previous section, additional effects are required
to obtain a static solution for the internal space. To ensure that the
extra dimensions do not shrink down to zero size,
 one can include the Casimir energy contribution due to the
various fields. Such a term is of the form $A/b^{4+D}$~[26],
 where $A$ is a constant.

This model was dimensionally reduced via the same procedure, and the effective
4-dimensional action was studied in the conformally transformed Einstein
frame in [25]. As was shown in [27], it is indeed the Pauli metric (in
the Einstein frame) which obeys the correct equation of motion for a
massless spin-2 graviton in four dimensions, as opposed to the metric in
the Jordan frame. In the Einstein frame, one has the further advantage that
scalar fields in the matter sector have canonical couplings to the metric,
and also standard kinetic terms in four dimensions~[25,27]. Apart from 
the inflaton, the present model has two more scalar fields (JBD type) in
four dimensions, one related to the internal scale factor, $\phi$, the other
coming from the ten dimensional JBD field $\Phi$.

The equations of motion were integrated to obtain exponentially expanding
solutions for the four dimensional scale factor $a(t)$, together with
compactifying solutions for the internal scale factor. Nucleation of true
vacuum bubbles proceed in a favorable manner, as in the models studied
earlier. At the end of the inflatonary phase transition, a radiation dominated
era is brought about by demanding the total effective cosmological constant
to vanish. With this condition, the potential for $\phi$ in the 
post-inflationary era is given by
\be
V(\phi) = c_1e^{-2\Phi + \phi} + c_2e^{-2\Phi + 8\phi/3} - c_3e^{-\Phi + 4\phi/3}
\en
where $c_1, c_2$ and $c_3$ are constants arising from combinations of 
parameters of this model~[25]. At the end of the inflationary phase
transition, the internal scale factor  undergoes oscillations about
the minimum of its potential with decreasing amplitude. Thus one is
subsequently led to a scenario wherein a stable internal manifold
with finite size occurs. The Einstein theory of gravitation describes
the cosmic evolution of the four dimensional spacetime in the post-inflationary
stage.

\section{Constraints on parameters of Kaluza-Klein models}

An unsatisfactory feature of higher dimensional models, as we have seen, is
the presence of several `free' parameters, such as the nonminimal coupling
parameter, higher dimensional JBD parameter, strength of the cosmological
term, etc. It is natural to expect that considerations for implementing
a successful EI scenario would impose restrictions on the values of these
parameters. The conditions for successful EI can be summarized~[28] as follows.

The primary requirement is that one needs to obtain enough inflation
$(H.x \ge e^{66})$, before the completion of the inflationary phase 
transition for solving the horizon and flatness problems. Secondly,a bubble
spectrum should ensue which agrees with the CMBR anisotropy $(\nabla T/T \sim
10^{-5})$ as reported by COBE~[29]. The quantum fluctuations of
the scalar field trapped in the local minimum of its potential should
generate suitable density perturbations that are compatible with the 
subsequent formation of large scale structure in the universe~[30]. Apart
from these requirements on the physics of the phase transition during the
inflationary era, one must recover a radiation dominated universe
described by the Einstein theory (or JBD theory with $\o > 500$) after the
end of inflation. One must also obtain the correct value for the four
dimensional gravitational constant.

All the above criteria for successful EI have been incorporated in case of
KK models described in this paper, and restrictions on their parameters
obtained in [31]. As we have seen here, in all these models the 4-dimensional
effective action contains a potential for the JBD field with a minimum
where it can settle down at the end of inflation. Hence, in essence, an
Einstein theory ensues at the end of inflation for the 4-dimensional
spacetime, whereas, the internal dimensions are compactified to a small
radius. Obtaining the correct value of $G$ leads to a constraint on a 
parameter of the higher dimensional action.

However, the stability of compactification cannot be ensured in case of the
minimal ten dimensional model, considered in Section 3. A generic feature
of such models is a suitable bubble spectrum  as demanded by CMBR anisotropy.
Hence, this requirement imposes rather weak constraints on the higher 
dimensional parameters. The most stringent condition is that of appropriate
density perturbations. This leads to severe restrictions on the parameters,
and it was observed in [31] that out of the models considered here, only
the ten dimensional JBD model is able to meet this requirement within a 
narrow range of allowed parameter space. 

\vskip 0.2in

\noindent
{\bf References}

\begin{description}

\item[[1]] See for instance, S.Weinberg, ``Gravitation and Cosmology'', 
J.Wiley and Sons
(1972).
\item[[2]] For example, see R.Brandenberger, Rev. Mod. Phys. {\bf 57} (1985) 1.
\item[[3]] See     A.D.Linde, ``Particle Physics and Inflationary Cosmology'',
Harwood Academic Publishers (1990).
(1990).
\item[[4]] A.H.Guth, Phys. Rev.  D {\bf 23} (1981) 347.
\item[[5]] A.H.Guth and E.J.Weinberg, Nucl. Phys. B {\bf 212} (1983) 321.
\item[[6]] A.Albrecht and P.Steinhardt, Phys. Rev. Lett. {\bf 48} (1982)
1220.
\item[[7]] A.D.Linde, Phys. Lett. B {\bf 129} (1983) 177.
\item[[8]] C. Mathiazhagan    and  V. B .Johri, Class. Quant. Grav. {\bf 1}
(1984) L 29;
        D.La and P.J.Steinhardt, Phys. Rev. Lett. {\bf 62} (1989) 376.
\item[[9]]  P.G.Bergmann, Int. J. Theor. Phys. {\bf 1} (1968) 25;
        R.V.Wagoner, Phys. Rev. D {\bf 1} (1970) 3209;
        K.Nordtvedt, Ap. J. {\bf 161} (1970) 1059.
\item[[10]]  P.Jordan, Z. Phys. {\bf 157} (1959) 112;
     C.Brans and R.H.Dicke, Phys. Rev. {\bf 124} (1961) 925.
\item[[11]] C.M.Will, Phys. Rep. {\bf 113} (1984) 345. 
\item[[12]]  S.Coleman, Phys. Rev. D {\bf 15} (1977) 2929;
        S.Coleman and F.Deluccia, {\it ibid.} {\bf 21} (1980) 3305. 
\item[[13]] D.La, P.J.Steinhardt and E.W.Bertschinger, Phys. Lett.
B {\bf 231} (1989) 231;
        E.J.Weinberg, Phys. Rev. D {\bf 40} (1989) 3950.   
\item[[14]] R.Holman, E.Kolb and Y.Wang, Phys.
Rev. Lett. {\bf 65} (1990) 17; 
A. Layock and
A.R.Liddle, Phys. Rev. D {\bf 49} (1994) 1827;  B. Chakraborty and 
T. R. Seshadri, Astroparticle Phys. {\bf 5} (1996) 35.
\item[[15]] N.Panchapakesan and S.K.Sethi, Int. J. Mod. Phys. A {\bf 7} (1992)
6665.        
\item[[16]] A.D.Linde, Phys. Lett. B {\bf 249} (1990) 18;
         J.D.Barrow and
K.Maeda, Nucl. Phys. B {\bf 341}  (1991) 294;
A.R.Liddle and D.Wands, Phys. Rev. D {\bf 45}
 (1992) 2665. 
\item[[17]] Th.Kaluza, {\it Sitzungsber. Preuss. Akad. Wiss. Phys. Math.
Klasse} {\bf 966} (1921) K 1;
        O.Klein, Nature {\bf 118} (1926) 516.
\item[[18]] ``Modern Kaluza-Klein theories'', eds. T. Appelquist, A. Chodos,
 P.G.O.Freund (Addison-Wesley 1987).
\item[[19]] A.Salam and J.Strathdee, Ann. Phys. {\bf 141} (1982) 163.
\item[[20]] S.Weinberg, in ``Inner Space / Outer Space'', eds. E.W.Kolb,
M.S. Turner, D.Lindley, K.Olive and D.Seckel (University of Chicago
Press 1986).
\item[[21]] R.Holman, E.W.Kolb, S.L.Vadas and Y.Wang, Phys. Rev. D
{\bf 43} (1991) 995.
\item[[22]] R.Holman, E.W.Kolb, S.L.Vadas, Y.Wang and E.J.Weinberg,
Phys. Lett. B {\bf 237} (1990) 37;
         R.Holman , E.W.Kolb , S.L.Vadas and Y.Wang, {\it ibid}.
{\bf 250} (1990) 24.
\item[[23]] K.Sunahara, M.Kasai and T.Futamase, Prog. Theor. Phys.
{\bf 83} (1990) 353.
\item[[24]] A.S.Majumdar and S.K.Sethi, Phys. Rev. D {\bf 46} (1992) 5315.
\item[[25]] A.S.Majumdar, T.R.Seshadri and S.K.Sethi, Phys. Lett. B {\bf 312}
(1993) 67.
\item[[26]]  S.Randjbar-Daemi, A.Salam and J.Strathdee, Phys. Lett.
B {\bf 135} (1984) 388;
         F.S.Accetta, M.Gleiser, R.Holman and E.W.Kolb, Nucl. Phys.
B {\bf 276} (1986) 501. 
\item[[27]] Y.M.Cho, Phys. Rev. Lett. {\bf 68} (1992) 3133.
\item[[28]] A.M.Green and A.R.Liddle, Phys. Rev. D {\bf 54} (1996) 2557.
\item[[29]] G.F.Smoot et al, Ap. J. {\bf 396} (1992) L 1.
\item[[30]] A.R.Liddle and D.H.Lyth, Phys. Lett. B
{\bf 291} (1992) 391;
 A.R.Liddle,
D.H.Lyth, R.K.Shaeffer, Q.Shafi and P.T.P.Viana,
Mon. Not. R. Astron. Soc. {\bf 281} (1996) 531.
\item[[31]] A.S.Majumdar Phys. Rev. D {\bf 55} (1997) 6092.

\end{description}

\end{document}